\magnification=\magstep1
\baselineskip=18pt

\def\sdp{$$\,\bigcirc\!\!\!\! s\,$$} 

\null

\centerline{World Spinors - Construction and Some Applications}
\vskip10pt
\centerline{Yuval Ne'eman
\footnote{$^{*}$}{Wolfson Distinguished Chair in Theoretical Physics}
\footnote{$^{\#}$}{Also on leave from Center for Particle Physics,
University of Texas, Austin, Texas 78712, USA}}
\centerline{Sackler Faculty of Exact Sciences, Tel-Aviv University}
\centerline{69978 Tel-Aviv, Israel}
\vskip10pt
\centerline{and}
\vskip10pt
\centerline{Djordje \v Sija\v cki
\footnote{$^{+}$}{Supported in part by the Science Foundation (Belgrade)}}
\centerline{Institute of Physics, P O Box 57, Belgrade, Yugoslavia}
\vskip40pt

\noindent ABSTRACT

The existence of a topological double-covering for the $GL(n,R)$ and
diffeomorphism groups is reviewed. These groups do not have
finite-dimensional faithful representations. An explicit construction
and the classification of all $\overline{SL}(n,R)$, $n=3,4$ unitary
irreducible representations is presented. Infinite-component spinorial
and tensorial $\overline{SL}(4,R)$ fields, "manifields", are introduced.
Particle content of the ladder manifields, as given by the 
$\overline{SL}(3,R)$ "little" group is determined.
The manifields are lifted to the corresponding world spinorial and 
tensorial manifields by making use of generalized infinite-component frame
fields. World manifields transform w.r.t. corresponding 
$\overline{Diff}(4,R)$ representations, that are constructed explicitly. 

\vfill
\eject
\vskip18pt
\noindent 1. INTRODUCTION 
\vskip9pt

Larry Biedenharn's contributions to physics span several of its 
subdisciplines, such as Atomic, Nuclear, or Particle Physics. The common 
denominator is his masterly handling of Group Theory, certainly a very 
powerful tool in these fields. One of the most beautiful examples of Larry's 
virtuoso performance is his solution [1] of the Racah problem: How 
does one characterize - with no degeneracies - the states in the unitary 
irreducible representations of $SU(3)$, when applying (e.g. as in harmonic 
oscillator models) the $SU(3)\to SO(3)$ reduction sequence, i.e. 
with the $SO(3)$ 3-dimensional vector spanning the same carrier space as 
the 3-dimensional defining representation of $SU(3)$. The problem caught 
the interest and imagination of the algebraic experts (including Racah 
himself), who worked on it, with the late Y. Lehrer-Ilamed for years. 
L.B.'s solution is "final" and also shows that there are no rational operator 
functions capable of fulfilling the task, while presenting the irrational 
functions which do. 

The authors of this article owe their mutual links, which produced their 
intensive twenty years' personal collaboration, to the fact that their careers 
intersected with Larry Biedenharn's, in the group theory context. The first 
author (YN) while at Caltech in 1963-65, happened to produce, in 
collaboration with M. Gell-Mann and with the late Yossef Dothan, a suggestion 
for a group-theoretical characterization of the hadron Regge sequences, till 
then charted phenomenologically, after the great resonance "explosion" in 
1960-61 [2]. The model also supplied an algebraic structural derivation, 
involving {\it gravitational quadrupoles}. This appeared rather surprising at 
the time, but has been explained by the present authors in recent years [3].
The algebraic Regge model [2], based on assignments to {\it ladder}-type 
infinite representations of the noncompact group $SL(3,R)$ (whose construction 
was also first given in [2]), with $\Delta J=2$ and for lowest spins 
$J_{0}=0,1,2$, also appeared to be extendable to nuclear physics. 
This is a part of Physics in which "quadrupolar" algebras (based on the 
harmonic oscillator $SU(3)$ degeneracy group) had been introduced by Elliott 
in 1958 [4]. 

YN first met Larry Biedenharn at the Coral Gables Conferences, and 
discussed these $SL(3,R)$ results and their possible relevance to Nuclei. 
Larry was interested and several years later (1970-73) indeed successfully 
applied the $SL(3,R)$ algebra to nuclei [5]. Indeed, one now has a good 
understanding [6] of the intertwining of the three different algebras 
($SL(3,R), SU(3), SO(3)\times T_(3)$) which can be generated by the 
commutators between angular momentum and quadrupole operators. 

And yet $SL(3,R)$ went "deeper". The question of {\it the existence of a
double covering} had already arisen in 1965, when the authors of [2] looked
for an $SL(3,R)$ assignment, to fit the fermionic Regge trajectories. This
had, however, remained unanswered. In 1969, when YN was next at Caltech for
a term, he initiated an algebraic study of the case, together with Dr D.W. 
Joseph, of the University of Nebraska, with whom he had collaborated in 1964
in a Kaluza-Klein approach to (flavor) $SU(3)$. The answer to the question
of the existence of a double-covering was indeed positive, there is such a
$\overline{SL}(3,R)$ group with only infinite unitary representations and  
one should thus have been able to utilize these unitary irreducible infinite
representations for fermionic sequences. However, an unexpected difficulty
suddenly emerged in this program, in the form of a {\it singularity}
occurring in the "ladder"-like representation whose lowest state is
$J_{0}=3/2$ (needed for the "most important" hadron resonance, Fermi's 
$(I=3/2,J=3/2)$). Note that there was no difficulty with $J_{0}=1/2$. David 
Joseph prepared a preprint for publication [7], but the enthusiasm for 
publication had waned for YN, as the answer appeared to fail for 
the most important physical case, the $I=3/2, J=3/2$. Joseph sent out his 
preprint, which was never published, as a result of a combination of 
referee difficulties and loss of enthusiasm. However, the preprint did  
trigger a renewal of interest in $SL(3,R)$ among the group theory fans, 
including in Larry's group at Duke University [5]. The difficulty with 
$J_{0}=3/2$ was first glossed over, but then resurfaced, with a contribution 
[8] from another group-theory virtuoso, the late V. Ogievetsky, who died in 
the same year 1995 as Larry Biedenharn, in a sports accident. 

Meanwhile, the second author (Dj. \v{S}.) had arrived in 1972 at Duke 
University, becoming engaged in a doctoral dissertation program. With 
the interest in $SL(3,R)$ as displayed in both particle and nuclear physics, 
it seemed worth investing a real effort in charting the entire system of 
representations, including those of the double-covering. D.S.'s results, 
published in 1975 [9], were extensive and "final", as emphasized several 
years later in a more mathematically oriented study [10]. Thus, when in 
1977, YN demonstrated [11, 12, 13] the relevance of these results to an 
issue in Gravity, namely the erroneous ruling-out of curved space spinors 
({\it world spinors}), it was natural that the two authors should converge 
in their interests - and the present collaboration was born.  

We now present the problem from that gravitational angle.

In the standard approach to General Relativity one starts with the group
of "general coordinate transformations" ($GCT$), i.e. the group of
diffeomorphisms Diff(R$^4$). The theory is set upon the principle of
general covariance. A unified description of both tensors and spinors
would require the existence of respectively tensorial and (double valued)
spinorial representations of the $GCT$ group. In other words one is
interested in the corresponding single-valued representations of the double
covering $\overline{GCT}$ of the $GCT$ group, since the topology of $GCT$
is given by the topology of its linear compact subgroup. It is well known
that the finite-dimensional representations of $\overline{GCT}$ are
characterized by the corresponding ones of the
$\overline{GL}(4,R)\supset\overline{SL}(4,R)$ group, and
$\overline{SL}(4,R)$ does not have {\it finite} spinorial representations.
However there are infinite-dimensional spinors of $\overline{SL}(4,R)$
which are the true "world" (holonomic) spinors [14]. There are two ways
to introduce finite spinors: i) One can make use of the nonlinear
representations of the $\overline{GCT}$ group, which are linear when
restricted to the Poincar\'e subgroup [15]. ii) One can introduce a
bundle of cotangent frames, i.e. a set of 1-forms $e^a$ (tetrads;
$a=0,..., 3$ the anholonomic indices) and define in this space an action
of a physically distinct local Lorentz group. Owing to this Lorentz group
one can introduce finite spinors, which behave as scalars w.r.t.
$\overline{GCT}$. The bundle of cotangent frames represents an additional
geometrical construction corresponding to the physical constraints of a
local gauge group of the Yang-Mills type, in which the gauge group is the
isotropy group of the space-time base manifold. One is now naturally led
to enlarge the local Lorentz group to the whole linear group
$\overline{GL}(4,R)$, and together with translations one 
obtains the affine group $\overline{GA}(4,R)$. The affine group translates 
and deforms the tetrads of the locally Minkowskian space-time [16], and 
provides one with either infinite-dimensional linear or finite-dimensional
nonlinear spinorial representations [17].

The existence and structure of spinors in a generic curved space have
been the subject of more confusion than most issues in mathematical
physics. True, to the algebraic topologist the problem appears to have
been answered long ago, with the realization that {\it the topology of a
noncompact Lie group follows that of its maximal compact subgroup}. This
perhaps is the reason for the low priority given by mathematicians, in the
case of the linear groups, to the study of the representations of their
double-covering, for instance [10]. 

The issue is an important one for the physicist, however, and we shall
make one more effort to clarify it. The physics literature contains two
common errors. For fifty years, it was wrongly believed that the
double-covering of $GL(n,R)$, which we shall denote
$\overline{GL}(n,R)$ does not exist. Almost every textbook in general
relativity theory, upon reaching the subject of spinors, contains a
sentence such as "... there are no representations of $GL(4,R)$, or
even 'representations up to a sign', which behave like spinors under
the Lorentz subgroup". Though the correct answer has been known since 1977 
[11, 12, 13], the same type of statement continues to appear in more recent 
texts. The present authors were much encouraged in their dealing with the 
issue of spinors in a curved space by the convergence of their interests 
in this matter with the investigation of Metric-Affine manifolds initiated 
by F.W. Hehl and his Cologne group [16]. The contents of a recent review 
[18] testify to the richness of the subject.  

An additional reason for the overall confusion concerns the unitarity of 
the relevant spinor representations. In dealing with noncompact groups,
it is customary to select infinite-dimensional unitary representations,
where the {\it particle-states} are concerned. For both tensor or
spinor {\it fields}, however, finite and nonunitary representations are
used (of $GL(4,R)$ and $SL(2,C)$ respectively). We showed that the 
correct answer for spinorial $\overline{GL}(4,R)$ fields consists in 
using the infinite unitary representations in a physical base in which
they become nonunitary [19].

In recent years, the unitary infinite-dimensional representations of
the double-cov\-erings $\overline{GL}(n,R)$ and $\overline{SL}(n,R)$ have
been classified and constructed for $n=3$ [9], $n=4$ [20], while the
case $n=2$ has been known for many years [21]. Field equations have
been constructed for such infinite-component fields, "manifields", within
Riemannian gravitational theory and for Einstein-Cartan gravity [22],
including the case of "world spinors" [14], and for affine [17, 23, 24] 
gravity. $\overline{SL}(4,R)$ manifields have also been used in
classifying the hadron spectrum [25, 26].

\vskip18pt
\noindent 2. EXISTENCE OF THE DOUBLE-COVERING $\overline{GL}(n,R)$ 
\vskip9pt

The basic results can be found in Ref. [27].
Let $g_{0}=k_{0}+a_{0}+n_{0}$ be an Iwasawa decomposition of a
semisimple Lie algebra $g_{0}$ over $R$. Let $G$ be any connected Lie
group with Lie algebra $g_{0}$, and let $K,A,N$ be the analytic
subgroups of $G$ with Lie algebras $k_{0}$,$a_{0}$ and $n_{0}$
respectively. The mapping
$(k,a,n)\rightarrow kan\quad (k\in K,a\in A,n\in N)$
is an analytic diffeomorphism of the product manifold $K\times A\times N$
onto $G$. The groups $A$ and $N$ are simply connected.
Any semisimple Lie group can be decomposed into the product of the 
maximal compact subgroup $K$, an Abelian group $A$ and a 
nilpotent group $N$. As a result of, only $K$ is not
guaranteed to be simply-connected. There exists a universal covering
group $\overline{K}_u$ of $K$, and thus also a universal covering of
$G$:
$$
\overline{G}_{u} \simeq \overline{K}_{u} \times  A \times  N. 
$$
For the group of diffeomorphisms, let $Diff(n,R)$ be the group of all
homeomorphisms $f$ or $R^{n}$ such that $f$ and $f^{-1}$ are of class
$C^{1}$.  In the neighborhood of the identity
$$
V_{r,\varepsilon} = \left\{ g\in Diff(n,R)\mid \left[g(x)-x\right] <
\varepsilon, [{\partial g_{i}\over \partial x_{k}}(x) - \delta_{k}^{i}]
< \varepsilon , \mid x \mid < r \quad i,k = 1,...,n \right\} 
$$
Stewart [28] proved the decomposition
$$
Diff(n,R) = GL(n,R) \times H \times R_{n} 
$$
where the subgroup $H$ is contractible to a point. As $O(n)$
is the compact subgroup of $GL(n,R)$, one finds that 
$O(n)$ is a deformation retract of $Diff(n,R)$.
Thus, there exists a universal covering of the Diffeomorphism group
$$
\overline{Diff}(n,R)_{u} \simeq 
\overline{GL}(n,R)_{u} \times H \times R_{n}. 
$$

Summing up, we note that both $SL(n,R)$ and on the other hand $GL(n,R)$
and $Diff(n,R)$ will all have double coverings, defined by
$\overline{SO}(n) \simeq Spin(n)$ and $\overline{O}(n) \simeq Pin(n)$ 
respectively, the double-coverings of the $SO(n)$ and $O(n)$ maximal compact 
subgroups.

\vskip18pt
\noindent 3. $\overline{SL}(3,R)$ AND $\overline{SL}(4,R)$ UNIRREPS
\vskip9pt

$SL(n,R)$ is the group of linear unimodular transformations in an
$n$-dimensional real vector space. The group is a simple and noncompact
Lie group. The space of the group parameters is not simply connected. The
maximal compact subgroup of $SL(n,R)$ is $SO(n)$. The double covering 
(the universal covering for $n > 2$) group of $SL(n,R)$ we
denote by $\overline{SL}(n,R)$. Its maximal compact subgroup is
$\overline{SO}(n)\simeq Spin(n)$, the covering group of $SO(n)$.
$$
\overline{SL}(n,R)/Z_2\simeq SL(n,R),\quad \overline{SO}(n)/Z_2\simeq SO(n).
$$
In order to present the explicit forms of the $\overline{SL}(n,R)$
generators, $n=3,4$, we first separate them according to compactness and
it is most convenient to take them in the spherical basis. We list a 
minimal set of commutation relations. The remaining ones can be obtained 
by means of the Jacobi identity.

The $\overline{SL}(3,R)$ generators are $J_0,\ J_\pm ,\ T_M ,\ M =0,\pm
1,\pm 2$. $J_0$ and $J_\pm$ generate the $SU(2)$ subgroup, while $T_M$
forms an $SU(2)$ second rank irreducible tensor operator. The commutation
relations are:
$$
\eqalign{&[J_0, J_\pm ]=\pm J_\pm ,\quad [J_+, J_-]=2J_0,\quad [J_0, T_M] 
= M T_M,\cr
&[J_\pm , T_M ] = \sqrt{6-M(M\pm 1)} T_{M\pm 1}\quad 
[T_{+2}, T_{-2}] = - 4J_0.\cr}
$$

The $\overline{SL}(4,R)$ generators are $J_0^{(i)},\ J_\pm^{(i)},\ 
Z_{pq},\ i=1,\ 2;\ p,\ q = 0,\ \pm 1$. $J_0^{(i)}$ and
$J_\pm^{(i)}$ generate the $SU(2)\otimes SU(2)$ subgroup, while
$Z_{pq}$ forms, w.r.t. $SU(2)\otimes SU(2)$, a $(1,1)$-irreducible
tensor operator. The commutation  relations are:
$$
\eqalign{&[J_0^{(i)}, J_\pm^{(j)}]=\pm\delta_{ij} J_\pm^{(i)},\quad
[J_+^{(i)}, J_-^{(j)}]=2\delta_{ij}J_0^{(i)},\quad [J_0^{(i)},
J_0^{(j)}]=0\cr 
&[J_0^{(1)}, Z_{pq}] = pZ_{pq},\quad [J_0^{(2)},
Z_{pq}]= qZ_{pq}\cr
&[J_\pm^{(1)}, Z_{pq}]=\sqrt{2-p (p\pm 1)}Z_{p\pm 1,q},\quad 
[J_\pm^{(2)}, Z_{pq}] = \sqrt{2-q(q\pm 1)} Z_{p,q\pm 1},\cr
&[Z_{+1,+1}, Z_{-1,-1}]=-(J_0^{(1)}+J_0^{(2)}).\cr}
$$

In order to analyze the representations, as well as to make use of them in
a gauge theory, it is convenient to have the matrix elements of the group 
generators. Also, in
this case the task of determining the scalar products of the unitary
representations is considerably simplified. The most general results are
obtained in the $\left|{{j}\atop {k\  m}}\right>$, $\left| {{j_{1}}\atop 
{k_{1}\ m_{1}}} \ {{j_{2}}\atop {k_{2}\ m_{2}}} \right>$ basis of the 
$SU(2)$, $SU(2)\otimes SU(2)$ representations respectively, 
$j,\ j_1,\ j_2=0,\ 1/2,\ 1\ ...$ The matrix elements of the compact 
generators are well known, and we list only the matrix elements of the 
noncompact generators [10, 20].

\noindent {\bf n=3}:
$$
\left< {j^\prime}\atop {k^\prime\ m^\prime}
\right| T_M \left| {j}\atop {k\ m}\right> = (-)^{j^\prime-m^\prime}
\left(\matrix{j^{\prime}&2&j\cr -m^{\prime}&M&m\cr}\right)
\left< {j{\prime}}\atop {k^{\prime}}\right\Vert T \left\Vert {j}\atop
{k}\right>, \quad M=0,\pm 1,\pm 2,
$$
where,
$$
\left< {j{\prime}}\atop {k^{\prime}}\right\Vert T \left\Vert {j}\atop
{k}\right>
= (-)^{j^{\prime}-k{^\prime}}\sqrt{(2j^{\prime}+1)(2j+1)} \Biggl\{
{-i\over\sqrt{6}}[2\sigma -j^\prime (j^\prime +1)+j(j+1)] 
\left(\matrix{j^{\prime}&2&j\cr -k^{\prime}&0&k\cr}\right) +
$$
$$
+i(\delta + k + 1)
\left(\matrix{j^{\prime}&2&j\cr -k^{\prime}&2&k\cr}\right) + i(\delta -k+1) 
\left(\matrix{j^{\prime}&2&j\cr -k^{\prime}&-2&k\cr}\right)  \Biggr\} ,
$$
$\sigma =a+b,\  \delta =a-b$.

\noindent {\bf n=4}:

$
\left< {{j_1^\prime}\atop {k_1^\prime \ m_1^\prime}}\ 
{{j_2^\prime}\atop {k_2^\prime \ m_2^\prime}}\right| 
Z_{pq}\left|{{j_1}\atop {k_1\ m_1}}\ {{j_2}\atop {k_2\ m_2}}\right>=
$
$$
=(-)^{j_1^\prime -m_1^\prime}(-)^{j_2^\prime-m_2^\prime} 
\left(\matrix{j_1^\prime &1&j_1\cr -m_1^\prime &p &m_1\cr}\right)
\left(\matrix{j_2^\prime &1&j_2\cr -m_2^\prime &q &m_2\cr}\right)
\left< {{j_1^\prime}\atop {k_1^\prime}}{{j_2^\prime}\atop {k_2^\prime}}
\right\Vert Z \left\Vert {{j_1}\atop {k_1}}{{j_2}\atop {k_2}}\right>,
$$
where,
$$
\left< {{j_1^\prime}\atop {k_1^\prime}}{{j_2^\prime}\atop {k_2^\prime}}
\right\Vert Z \left\Vert {{j_1}\atop {k_1}}{{j_2}\atop {k_2}}\right> =
(-)^{j_1^\prime-k_1^\prime}(-)^{j_2^\prime -k_2^\prime}{i\over 2}
\sqrt{(2j_1^\prime +1)(2j_2^\prime +1)(2j_1+1)(2j_2+1)}\times
$$
$$
\times\Biggl\{ [e+4-j_1^\prime (j_1^\prime +1)+j_1(j_1+1)-j_2^\prime
(j_2^\prime +1)+j_2(j_2+1)]
\left(\matrix{j_1^\prime &1&j_1\cr -k_1^\prime &0 &mk_1\cr}\right)
\left(\matrix{j_2^\prime &1&j_2\cr -k_2^\prime &0 &k_2\cr}\right)
$$
$$
-(c+k_1-k_2)\left(\matrix{j_1^\prime &1&j_1\cr -k_1^\prime &1 &mk_1\cr}\right)
\left(\matrix{j_2^\prime &1&j_2\cr -k_2^\prime &-1 &k_2\cr}\right)
$$
$$
-(c-k_1+k_2)\left(\matrix{j_1^\prime &1&j_1\cr -k_1^\prime &-1 &mk_1\cr}\right)
\left(\matrix{j_2^\prime &1&j_2\cr -k_2^\prime &1 &k_2\cr}\right)
$$
$$
+(d+k_1+k_2)\left(\matrix{j_1^\prime &1&j_1\cr -k_1^\prime &1 &mk_1\cr}\right)
\left(\matrix{j_2^\prime &1&j_2\cr -k_2^\prime &1 &k_2\cr}\right)
$$
$$
+(d-k_1-k_2)\left(\matrix{j_1^\prime &1&j_1\cr -k_1^\prime &-1 &mk_1\cr}\right)
\left(\matrix{j_2^\prime &1&j_2\cr -k_2^\prime &-1 &k_2\cr}\right)\Biggr\},
$$
$e=c-a-b,\ d=a-b.$

\noindent The representation labels $\sigma$, $\delta$ (for $n=3$); and $c$,
$d$, $e$ (for $n=4$) are arbitrary complex numbers and are determined 
from the representation space scalar product's unitarity and from the group 
generators' hermiticity requirements.

We now list all unitary irreducible representation labels and the 
$\overline{SO}(3,R)$ subgroup labels of the $\overline{SL}(3,R)$ group [9].

\noindent Principal series: $\sigma_1=\delta_1=0,\quad
\sigma_2,\delta_2\in R$

\hskip1.5cm $(\varepsilon ,\varepsilon^\prime)=(+1,+1):\  \{ j\} =\{ 
0^1,2^2,3^1,4^3,5^2,...\}$

\hskip1.5cm $(\varepsilon ,\varepsilon^\prime)=(+1,-1),\ (-1,\pm 1):\  
\{ j\} =\{1^1,2^1,3^2,4^2,5^3,...\} ,\ 
\{ {1\over 2}^1,{3\over 2}^2,{5\over 2}^3,...\}.$

\noindent Supplementary series: $\sigma_1=\delta_2=0,\quad \sigma_2\in R$

\hskip1.5cm $0 < \delta_1 < 1, \quad (\varepsilon ,\varepsilon^\prime) = 
(+1,+1): \ \{ j\} =\{ 0^1,2^2,3^1,4^3,5^2,...\}$

\hskip3.6cm $(\varepsilon ,\varepsilon^\prime)=(+1,-1): \ \{ j\} = 
\{1^1,2^1,3^2,4^2,5^3,...\}$

\hskip1.5cm $0 < \delta_1 \leq {1\over 2}, \quad \{ j\} = \{ {1\over 
2}^1,{3\over 2}^2,{5\over 2}^3,...\}$

\noindent Discrete series: $\sigma_1=\delta_2=0, \ \sigma_2\in R, \ 
\delta_1=1-\underline{j};\ \underline{j}= {3\over 2},2,{5\over 2},3,...$

\hskip1.7cm $\{ j\} =\{ \underline{j}^1, (\underline{j}+1)^1,
(\underline{j}+2)^2, (\underline{j}+3)^2, (\underline{j}+4)^3,...\}$

\noindent Multiplicity free (ladder) series: $\sigma_1=\delta_2=0,\  
\delta_1=1,$

\hskip1.7cm $\sigma_2\in R,\quad \{ j\} =\{ 0,2,4,...\} ,\  \{ j\} =\{ 
1,3,5,...\}$

\hskip1.7cm $\sigma_2=0,\quad \{ j\} =\{ {1\over 2},{5\over 2},{9\over
2},...\}$.

For the general case of the $\overline{SL}(4,R)$ unirreps we present here
only the labels. For the general (multiplicity non free) case, we have [20]

A) $e_1=0,\  e_2\in R,$

B$_1$) $d_1=0,\  d_2\in R,$

B$_2$) $d_1=\underline{k}_1+\underline{k}_2,\  d_2=0; \quad
\underline{k}_1+\underline{k}_2={1\over 2},1,{3\over 2},...,$

B$_3$) $0 < d_1 < 1,\  d_2=0; \quad k_1+k_2=0,\pm 2,\pm 4,...,$

B$_4$) $0 < d_1 < {1\over 2},\  d_2=0; \quad k_1+k_2\equiv {1\over 2}
(mod2)\quad {\hbox {or}}\quad {3\over 2}(mod2),$

C$_1$) $c_1=0,\  c_2\in R,$

C$_2$) $c_1=\underline{k_1}-\underline{k_2},\  c_2=0;\quad
\underline{k_1}-\underline{k_2}={1\over 2},1{3\over 2},...,$

C$_3$) $0 < c_1 < 1,\  c_2=0;\quad k_1-k_2=0,\pm 2,\pm 4,...,$

C$_4$) $0 < c_1 < {1\over 2},\  c_2=0;\quad k_1-k_2={1\over 2}
(mod2)\quad {\hbox {or}}\quad {3\over 2}(mod2).$

\noindent Any combination of (A) with one (B) and one (C) determines a
series of $\overline{SL}(4,R)$ unirreps. For these series $j_1\geq|k_1|,
\  j_2\geq|k_2|$. There are four series of multiplicity free
$\overline{SL}(4,R)$ unirreps [19].

\noindent Principal series: $e_1=0,\  e_2\in R;\quad j_1+j_2\equiv 
0(mod2)\quad {\hbox {or}}\quad 1(mod2),$

\noindent Supplementary series: $0 < e_1 < 1,\  e_2=0;\quad j_1+j_2\equiv
1(mod2),$

\noindent Discrete series: $e_1=1-\underline{j},\  e_2=0;\quad
\underline{j}={1\over 2},1,{3\over 2},...,|j_1-j_2| \geq\underline{j},$

\noindent Ladder series: $e_1=0,\  e_2\in R;\  j_1=j_2=j,\  \{ j\}=\{
0,1,3,...\} , \{ j\} =\{ {1\over 2},{3\over 2},{5\over 2}...\} .$

\vskip18pt
\noindent 4. $\overline{GA}(4,R)$ OR $\overline{SA}(4,R)$ MATTER FIELDS.
\vskip9pt

The general affine group $\overline{GA}(4,R) = T_{4}\sdp 
\overline{GL}(4,R)$, is a semidirect product of translations and
$\overline{GL}(4,R)$, the general linear group, generated by $Q_{ab}$.
Here $\overline{GL}(4,R) = R_{+} \otimes\overline{SL}(4,R) \supset R_{+}
\otimes \overline{SO}(1,3)$, where $R_{+}$ is the dilation subgroup. 
The antisymmetric
operators $Q_{[ab]} = {1\over 2}(Q_{ab}-Q_{ba})$ generate the Lorentz
subgroup $\overline{SO}(1,3)$, the symmetric traceless operators
(shears) $Q_{(ab)} = {1\over 2}(Q_{ab}+Q_{ba}) - 
{1\over 4}g_{ab}Q_{c}^{\ c}$ 
generate the proper $4$-volume-preserving deformations while the
trace $Q=Q^{a}_{\ a}$ generates scale-invariance $R_{+}$. $Q_{[ab]}$
and $Q_{(ab)}$ generate together the $\overline{SL}(4,R)$ group.

The $\overline{SA}(4,R)$ unirreps [19, 20] are induced from the
corresponding little group unirreps. The little group turns out to be
$\overline{SA}(3,R)^{\sim} = T_{3}^{\sim}\sdp\overline{SL}(3,R)$,
and thus we have the following nontrivial possibilities:

(i) $T_{3}^{\sim}$ is represented trivially, and the corresponding
states are described by the $\overline{SL}(3,R)$ unirreps, which are
infinite-dimensional owing to the $\overline{SL}(3,R)$
noncompactness. The corresponding $\overline{SL}(4,R)$ matter fields
are therefore necessarily infinite-dimensional and when reduced with
respect to the $\overline{SL}(3,R)$ subgroup should transform with
respect to its unirreps.

(ii) The little group $\overline{SA}(3,R)^{\sim}$ is represented
nontrivially, and we find the states which are characterized
"effectively" by three real numbers in addition to the
$\overline{SA}(2,R)^{\sim}$ unirreps.

(iii) For quarks or leptons, we make use of the $\overline{GA}(4,R)$ 
nonlinear representations which are realized through metric $g_{ab}$.
The stability subgroup is $SL(2,C)$, and the representations are linear
for the Poincar\'e subgroup.

Had the whole $\overline{SL}(4,R)$ been represented unitarily, the
Lorentz boost generators would have a hermitian intrinsic part; as a
result, when boosting a particle, one would obtain a particle with a 
different spin, i.e. another particle - contrary to experience. 
There exists however a remarkable inner {\it deunitarizing} 
automorphism ${\cal A}$ [19], which leaves the
$R_{+}\otimes\overline{SL}(3,R)$ subgroup intact, and which maps the
$Q_{(0k)}$, $Q_{[0,k]}$ generators into $iQ_{[0k]}$, $iQ_{(0k)}$
respectively $(k=1,2,3)$. The deunitarizing automorphism allows us to
start with the unitary representations of the $\overline{SL}(4,R)$
subgroup, and upon its application, to identify the finite (unitary)
representations of the abstract $\overline{SO}(4,R)$ compact subgroup
with nonunitary representations of the physical Lorentz group, while
the infinite (unitary) representations of the abstract
$\overline{SO}(1,3)$ group now represent (non-unitarily) the compact
$\overline{SO}(4)/\overline{SO}(3)$ generators. The non-hermiticity of
the intrinsic boost operators cancels their "intrinsic" physical action
precisely as in finite tensors or spinors, the boosts thus acting
kinetically only.  In this way, we avoid a disease common to
infinite-component wave equations.

Let us denote a
generic $\overline{SL}(4,R)$ unirrep by $D(c,d,e;(j_{1},j_{2}))$ where
$c$, $d$, $e$ are the representation labels, and $(j_{1},j_{2})$ denote
the lowest $\overline{SO}(4) = SU(2)\otimes SU(2)$ representation
contained in the given $\overline{SL}(4,R)$ representation.

For the $\overline{SL}(4,R)$ tensorial field we take an
infinite-component field $\Phi$ which transforms with respect to an
${\cal A}$-deunitarized unirrep belonging to the principal series of
representations $D_{SL(4,R)}^{pr}(c_{2},d_{2},e_{2},(00))$, $c_{2}$,
$d_{2}$, $e_{2}$ $\in R$. The manifield $\Phi$ obeys a
Klein-Gordon-like equation
$$ 
\big( g^{ab}\partial_{a}\partial_{b} + M^{2} \big) \Phi (x)=0. 
$$

For the $\overline{SL}(4,R)$ spinorial fields we take an
infinite-component field $\Psi$ which transforms with respect to an
${\cal A}$ deunitarized unirrep belonging to the principal series of
representations:
$D_{\overline{SL}(4,R)}^{pr}(c_{2},d_{2},e_{2};({1\over 2},0)) \oplus
D_{\overline{SL}(4,R)}^{pr}(c_{2},d_{2},e_{2};(0,{1\over 2}))$,
$c_{2},d_{2},e_{2}\in R$ while $({1\over 2},0)$ and $(0,{1\over 2})$
denote parity-conjugated spinorial representations. The manifield
$\Psi$ satisfies a Dirac-like equation
$$ 
\big( ig^{ab}\chi_{a}\partial_{b} - M \big) \Psi (x) = 0, 
$$
where $\chi_{a}$ is an $\overline{SL}(4,R)$ four-vector acting in the
space of our spinorial manifield. We construct $\chi_{a}$ in the
following way: first we embed $\overline{SL}(4,R)$ into
$\overline{SL}(5,R)$, and then select a pair of (mutually conjugate)
principal series representations which contain in the
$\overline{SL}(4,R)$ reduction our spinorial representations. Let the
$\overline{SL}(5,R)$ generators be $Q_{\hat{a}\hat{b}}$,
$\hat{a},\hat{b}=0,1,2,3,5$. We define $\chi_{a}=Q_{[5a]},a=0,1,2,3$
and thus arrive at the sought-for $\overline{SL}(4,R)$ four-vector.

\vskip18pt
\noindent 5. $\overline{SL}(3,R)$ CONTENT OF THE $\overline{SL}(4,R)$ 
LADDER REPRESENTATIONS
\vskip9pt

In order to study the  $\overline{SL}(3,R)$ irreducible representation
content of the $\overline{SL}(4,R)$ irreducible representations, it is 
convenient to define the following set of $\overline{SL}(4,R)$ algebra
 generators:
The compact generators are ($p, q, r = 0, \pm 1$)
$$
J_p = J^{(1)}_p + J^{(2)}_p ,\quad  
N_p = ({{-p}\over \sqrt{2}})^{|p|}(J^{(1)}_p - J^{(2)}_p) , \quad p=0,\pm 1 ,
$$
while the three noncompact $\overline{SO}(3)$ irreducible tensor operators
read
$$
Z^{(2)}_p = <2p|11qr> Z_{qr}, \quad
Z^{(1)}_p = <1p|11qr> Z_{qr}, \quad
Z^{(0)}_0 = <00|11qr> Z_{qr}. \quad
$$
The noncompact generators of the $\overline{SL}(3,R)$ algebra are 
$T_p = 2 Z^{(2)}_p$, while the boost generators are given by $K_p = 
i\sqrt{2} Z^{(1)}_p$. Moreover, in order to simplify the evaluation of
the relevant matrix elements, it is convenient to introduce the operator
$$
S = \sqrt{3} Z^{(0)}_0 ,
$$ 
that commutes with the entire  $\overline{SL}(3,R)$
group. 

The quantum numbers of the  $\overline{SL}(4,R)$ irreducible representation
decomposition w.r.t. its  $\overline{SL}(3,R)$ subgroup are determined by
the 
$$
\overline{SL}(4,R) \ \supset  \ R_{+}\otimes \overline{SL}(3,R)
\ \supset \ R_{+}\otimes \overline{SO}(3)
$$
group chain. The invariant subspaces of the $R_{+}$ subgroup generator $S$
determine the  $\overline{SL}(3,R)$ subgroup invariant subspaces as well - 
the nontrivial question is to determine whether these  $\overline{SL}(3,R)$ 
subspaces are irreducible or not, and finally to determine their
multiplicity. As for the irreducibility question, one can make use of the
$\overline{SL}(3,R)$ invariant, Casimir, operators. 

We will restrict ourselves, to the case of 
the ladder $\overline{SL}(4,R)$ irreducible representations and consider 
their decomposition as given by the 
above subgroup chain. First of all, one can prove that invariant 
eigen subspaces of the $R_{+}$ generator $S$, characterized by fixed $J$, 
$M$ quantum numbers of the $\overline{SO}(3)$ subgroup, are nondegenerate.
One can prove this statement by showing that all vectors with the same 
quantum numbers span one-dimensional subspaces. 

In the case of unitary irreducible representations of the $\overline{SL}(4,R)$
group, $S$ has to be represented by a Hermitian operator in the Hilbert space
and its eigenvalues are real numbers. Due to the fact that the set of $J$
quantum number values is unlimited (in contradistinction to the finite 
representation case), there are no constraint on the $S$ eigenvalues
whatsoever. Indeed in each invariant subspace the eigenvalues of $S$, say
$\alpha$, are arbitrary real numbers: $S |\ > = \alpha |\ >$, $\alpha \in R$.

Owing to the fact that the $\overline{SL}(3,R)$ group generators $T_p$ connect
the ladder representation states with  $\Delta J = \pm 2$, the $S$ invariant
subspaces of given $\alpha$ split into those of even and odd $J$ values. The
$\overline{SL}(3,R)$ Casimir operators, $C_2 = J\cdot J - {1\over 2}T\cdot T$
and $C_3 = J\cdot T\cdot J + {1\over 3} T\cdot T\cdot T$,  yield the 
following constraints [29] on the $\overline{SL}(3,R)$ and $\overline{SL}(4,R)$
representation labels ($\alpha , \sigma_2 , e_2 \in R$)
$$
\sigma_1 = 0 \quad\quad \sigma_2 = \alpha - 3e_2
$$

Finally, one finds that the ladder $\overline{SL}(4,R)$ unitary irreducible
representations decompose w.r.t the $R_{+} \otimes 
\overline{SL}(3,R)$ subgroup  representations according to the following
formula:
$$
D^{ladd}_{\overline{SL}(4,R)}(j;0,e_2) \quad \supset
$$
$$
\int^{\oplus} d\alpha
\{ [D_{R_{+}}(\alpha )
\otimes D^{ladd}_{\overline{SL}(3,R)}(0;0,\alpha -3e_2)]
\oplus
[D_{R_{+}}(\alpha )
\otimes D^{ladd}_{\overline{SL}(3,R)}(1;0,\alpha -3e_2)]\} ,
$$
where, $j = 0, {1\over 2}$ and $e_2 \in R$. Thus, to conclude, the ladder
unitary irreducible representations of the $\overline{SL}(4,R)$ group 
decompose into a direct integral of the $\overline{SL}(3,R)$ group 
ladder unitary irreducible representations.

\vskip18pt
\noindent 6. ANHOLONOMIC AND HOLONOMIC INDICES IN GRAVITY, 
WORLD SPINORS
\vskip9pt

Technically, it was the unembeddability of finite $\overline{SO}(1,3)$
spinors in finite (i.e. tensor) representations of $\overline{SL}(4,R)$
that required the 1929 introduction (by Hermann Weyl and by Fock and Ivanenko)
of the tetrad frames $e^a$ for curved space-time,
$$
e^a = e^a_\mu dx^\mu\ ,\quad e^a_\mu(\bar x)\equiv\bigg(\partial\xi^a(x)
/\partial x^\mu\bigg)_{x=\bar x}\ ,
$$
with the contraction
$$
e^ah_b = \delta^a_b\ ,\quad h_b = h_b^\mu(x)\partial_\mu\ .
$$
$\xi^a_{\bar x}$ is a set of coordinate axes erected at $x = \bar x$,
locally inertial there. Gravity then involves two invariance groups:
the anholonomic (tangent frame) group, here $\cal L$ and the covariance
group $\overline{Diff}(4,R)$. To achieve the overall
 transition to a local
tangent frame, we apply tetrads to the indices of a world-tensor
$$
\phi^{\mu\nu\cdots}(x)\to\phi^{ab\cdots}(x) =
e_\mu^a(x)e^b_\nu(x)\cdots\phi^{\mu\nu\cdots}(x)\ .
$$
The tetrad indices are contracted through the Minkowski metric
$\eta_{ab}$, while for world tensor indices this is achieved by the
metric $g_{\mu\nu}(x)$. The two are connected via
$$
\phi^a(x)\eta_{ab}\phi^b(x) = 
\phi^\mu(x)e^a_\mu(x)\eta_{ab}e^b_\nu(x)\phi^\nu(x)
= \phi^\mu(x)g_{\mu\nu}(x)\phi^\nu(x)\ .
$$
Note that the role of $\eta_{ab}$ is fulfilled in the finite Dirac
algebra by $\beta = \gamma^0$, for the spinor components.

The Principle of Equivalence is fulfilled for $\phi^{def\cdots}$ by
the following transition from flat to curved space ($\Lambda^c_b$ is
a numerical matrix representation of the $\overline{SO}(1,3)$ generators
on the $\phi^{def\cdots}$ basis)
$$
\partial_\mu\phi\to D_a\phi =
h_a^\mu(\partial_\mu-\omega_\mu^b{}_c\Lambda^c_b)\phi\ ,
$$
and in the opposite direction,
$D_a\phi\to\partial_\mu\phi\ ,\quad e^a_\mu\to\delta^a_\mu\ ,\quad
h_a^\mu\to\delta^\mu_a\ ,\quad\omega_\mu{}^{bc}\to0\ .$
At the same time, for world-tensor fields $\phi^{k\lambda\nu}$
$$
\partial_\mu\phi\to D_\mu\phi =
[\partial_\mu-\Gamma_\mu{}^\rho{}_\sigma(\Sigma_\rho^\sigma)]\phi\ ,
$$
where $\Sigma_\rho^\sigma$ is a numerical matrix representation of
the $\overline{SL}(4,R)$ generators on the $\phi^{k\lambda\nu\cdots}$ basis. 

What is special about the manifields $\Phi$ and $\Psi$
is that they do not have to be segregated in the local frame.
The unirreps of $\overline{SL}(4,R)$  support $\overline{Diff}(4,R)$  
and can thus be treated
holonomically. Mickelsson [30] has constructed an equation for a
holonomic (and non multiplicity-free) spinor in affine gravity, where
the flat limit does not hold, i.e. the extinction of the gravitational
field leaves a residual global $\overline{SL}(4,R)$  invariance and
thereby violates the Principle of Equivalence. However, this might fit
in a model in which the Lorentz group would emerge as the symmetry of
flat space-time after a further (spontaneous) symmetry breakdown [24]. 

To consider world spinors in ordinary riemannian Einstein
gravity, we denote by $\Psi^M(x), M, N = 1,2\cdots,\infty,$ the
$M$-component of the holonomic manifield, carrying a realization of
$\overline{Diff}(4,R)$, the covering group of general coordinate
transformations. In the local (anholonomic) frame, such a field obeys
the Lorentz invariant equation, i.e. its components $\Psi^U(x),
U, W = 1,2,\cdots,\infty,$ correspond to the reduction of the
representation $D^{\hbox{disc}}({1\over2},0)\oplus 
D^{\hbox{disc}}(0,{1\over2})$ of $\overline{SL}(4,R)$ 
over the infinite set of
representations of the compact sub-group $\overline{SO}(4)$,
representing here non-unitary finite representations of 
$\overline{SO}(1,3)$. We now define a pseudo-frame $E^U{}_M(x)$ s.t.
$$
\Psi^U(x) = E^U{}_M(x)\psi^M(x)\ .
$$
The $E^U{}_M(x)$ and their inverses $H^M{}_U(x)$ are thus infinite
matrices related to the quotient $\overline{Diff}(4,R)/ 
\overline{SL}(4,R)$ . Their transformation properties are
$$
\delta E^U{}_M(x) = -{1\over2}i\epsilon_b{}^a(x) 
\{\Lambda_a{}^b\}^U_VE^V{}_M(x)
+ \partial_\mu\xi^\rho\{\Sigma^\mu{}_\rho\}^N_ME^U{}_N(x) .
$$

Denoting by $B$ the constant $\gamma^0$-like matrix in the $X_\mu$ 
set in the manifield wave equation we have

$(\Psi^+(x))^U\{B\}_{UV}\Psi^V(x) \quad\to $
$$
(\Psi^+(x))^ME_M{}^U(x)
\{B\}_{uv}E_N{}^V(x)\Psi^N(x)
= (\Psi^+(x))^MG_{MN}(x)\Psi^N(x)\,
$$
where $G_{MN}(x)$ is a functional of the gravitational field 
realizing the metric on the world-spinor components.
The induced Riemannian condition, yields
$$
D_\mu E^U{}_M(x) = 0\ ,\quad D_\mu G_{MN}(x) = 0\ .
$$
In the absence of other spinor fields, the above equation involves the
Christoffel connection only,
$$
\partial_\mu G_{MN} - 
\Gamma_\mu{}^\rho{}_\sigma\{\Sigma_\rho{}^\sigma\}^P_MG_{PN}
- \Gamma_\mu{}^\rho{}_\sigma\{\Sigma_\rho{}^\sigma\}^P_NG_{MP} = 0\ ,
$$
which can be solved for $G_{MN}$, knowing $\Gamma$ and $\Sigma$.

The pseudo-frame $E^U{}_M$ can be realized geometrically in an
associate vector bundle over the bundle of linear frames.
$E^U{}_Md\Psi^M$ is a frame on the fiber.

We now consider the
infinite-dimensional representations (unirreps) of the double
covering $\overline{Diff}(4,R)$ of the group of analytic diffeomorphisms.
There is a rather elegant and economic method for this construction, 
which makes use of the pseudo-frames $E^U{}_M(x)$ and of the knowledge of 
the $\overline{{SL}}(4,R)$ unirreps.

The holonomic form of the $\overline{SO}(1,3)$ generators is given,
for an arbitrary infinite-dimensional representation, by $(H = E^{-1})$
$$
(M^a{}_b)^N{}_L(x) = H^N{}_U(x)(M^a{}_b)^U{}_VE^V{}_L(x)\ .
$$
In order to have a correct particle physics interpretation, we take
for the  $\overline{SO}(1,3)$ an infinite direct sum of
finite-dimensional non-unitary representations as explained. The
corresponding holonomic Lorentz-covariant matter field transforms
infinitesimally as follows:
$$
\delta\Psi^N(x) =
i\{\xi^\mu[\delta^N_L\partial_\mu + H^N{}_U(x)\partial_\mu E^U{}_L(x)] -
{1\over2}
\epsilon^b{}_aH^N{}_U(x)(M^a{}_b)^U{}_VE^V{}_L(x)\}\Psi^L(x)\ .
$$
An $\overline{SO}(1,3)$  infinitely reducible representation, which in 
its turn furnishes a basis for a $\overline{SL}(4,R)$  unirrep, can now
be lifted to a
$\overline{Diff}(4,R)$  representation. The corresponding holonomic
spinor/tensor manifields fit ordinary general relativity over a
riemannian space-time. A generalization to the metric-affine theory,
or even to the full general-affine theory, is rather straightforward.
The anholonomic Lorentz generators are substituted by the
$\overline{GL}(4,R)$ ones:
$$
Q^a{}_b = {1\over2}(M^a{}_b+T^a{}_b+{1\over2}\delta^a_bD)\ ,\quad a,b
= 0,1,2,3\ ,
$$
where $T^a{}_b$ are the four-volume-preserving shear-like generators,
while $D$ generates the dilation group. The holonomic version of
these generators, for an arbitrary unirrep, is given by
$$
(Q^a{}_b)^M{}_N(x) = H^M{}_U(x)(Q^a{}_b)^U{}_VE^V{}_N(x)\ .
$$
The transformation properties of a holonomic spinor/tensor manifield
are given as follows
$$
\delta\Psi^M(x) = i\{\xi^\mu[\delta^M_N\partial_\mu+H^M{}_U(x)\partial_\mu
E^U{}_N(x)]-\alpha^b_aH^M{}_U(x)(Q^a{}_b)^U{}_VE^V{}_N(x)\}\Psi^N(x)\ ,
$$
where $\alpha^b{}_a$ are $\overline{SL}(4,R)$ parameters. The
pseudo-frame under $U$ runs here over a basis of an
$\overline{SL}(4,R)$ unirrep. The resulting manifields 
transform with respect to $\overline{Diff}(4,R)$  according to the
representation generated by the operators $(Q^a{}_b)^M{}_N(x)$. 

An explicit construction of the $\overline{Diff}(4,R)$ unirreps requires,
a knowledge of the $\overline{SL}(4,R)$ unirreps.

If we consider
$$
\delta\Psi^M(x) =
i\xi^\mu\{\delta^M_N\partial_\mu+H^M{}_U(x)\partial_\mu E^U{}_N(x) - 
i\alpha^b{}_a
\partial_\mu[H^M{}_U(x)(Q^a{}_b)^U{}_VE^V{}_N(x)]\}\Psi^N(x)\ ,
$$
and make an expansion of the pseudo-frames in a power series of the
coordinates $x^\nu$, we obtain the corresponding representation of
the (infinite) Ogievetsky algebra, defined in the space of
manifield components. This algebra is generated by $\{P_\mu,F^{\nu_{1}, 
\nu_{2}\cdots\nu_{n}}_{\mu}
\mid n = 1,2,\cdots,\infty\}$ and the intrinsic part $F$ of these
generators is given by [31]
$$
\hat F^{\nu_{1},\nu_{2}\cdots\nu_{n}}_{\mu} = \partial_\rho(x^{\nu_{1}}
x^{\nu_{2}}\cdots x^{\nu_{n}})h^\rho{}_ae^b{}_\mu(Q^a{}_b)^U{}_V\ .
$$
Substituting here the generator matrix elements $(Q^a{}_b)^U{}_V$
of an $\overline{\hbox{SL}}$(4,R) unirrep we obtain the matrix
elements of the Ogievetsky algebra for the corresponding algebraic
representation of the $\overline{Diff}(4,R)$ group.

We close this review with a remark about possible future new applications  
of world spinors. Should the Quantum Superstring indeed "take over" 
as the fundamental theory of (Quantum) Gravity, it seems that the 
geometry beyond the Planck energy might well be nonriemannian. The 
structure of string theory already involves infinite linear representations, 
those of $Diff(R^{2})$. With the recent explosion in "dual" 
systems, in which superstrings become just one special case of "extendons" 
("p-branes") of dimensionalities $p\leq 6$ (in $D=11$, for instance), 
affine (or metric-affine) constructions might become the most convenient 
tool in dealing with systems supporting the action of $Diff(R^{p})$ [32].

\vskip18pt
\noindent {\it References:}
\vskip9pt

\item{1.}{L.C. Biedenharn, M.A. Locke and J.D. Louck, "The Canonical 
Resolution of the Multiplicity Problem for SU(3)", in {\it Proc. 4th Int. 
Coll. on Group Theoretical Methods in Physics}, Nijmegen (The Netherlands) 
1975, A. Janner, T. Janssen and M. Boon eds., Lecture Notes in Physics 
{\bf 50} (Springer Verlag, Berlin, Heidelberg, New York 1976), pp. 395-403.}
\item{2.}{Y. Dothan, M. Gell-Mann and Y. Ne'eman, {\it Phys. Lett.} {\bf 17} 
(1965) 148.}
\item{3.}{Dj. \v Sija\v cki and Y. Ne'eman, {\it Phys. Lett.} {\bf B 247} 
(1990) 571.} 
\item{4.}{J.P. Elliott, {\it Proc. Roy. Soc.} {\bf A 245} (1958) 128, 562.}
\item{5.}{L. Weaver and L.C. Biedenharn, {\it Phys. Lett.} {\bf B 32} 
(1970) 326.}
\item{6.}{J.P. Draayer and K.J. Weeks, {\it Phys. Rev. Lett.} {\bf 51} 
(1983) 1422.} 
\item{7.}{D.W. Joseph, U. of Nebraska preprint (unpublished) 1969.}
\item{8.}{V.I. Ogievetsky and E. Sokatchev, {\it Teor. Mat. Fiz.} {\bf 23} 
(1975) 462.}
\item{9.}{Dj. \v Sija\v cki, {\it J. Math. Phys.} {\bf 16} (1975) 298.}
\item{10.}{B. Speh, {\it Math. Ann.} {\bf 258} (1981) 113.}
\item{11.}{Y. Ne'eman, in {\it GR8} (Proc. 8th Int. Conf. on Gen. Rel. and 
Grav.) M.A. McKiernan, ed., Un. of Waterloo pub. (Waterloo, Canada, 1977) 269.}
\item{12.}{Y. Ne'eman, {\it Proc. Nat. Acad. Sci. USA} {\bf 74} (1977) 4157.}
\item{13.}{Y. Ne'eman, {\it Ann. Inst. H. Poincar\'e} {\bf A 28} (1978) 369.}   
\item{14.}{Y. Ne'eman and Dj. \v Sija\v cki, {\it Phys. Lett.} {\bf B 157} 
(1985) 267.}
\item{15.}{V.I. Ogievetsky and I.V. Polubarinov, {\it Soviet Phys. JETP} 
{\bf 21} (1965) 1093.}
\item{16.}{F.W. Hehl, G.D. Kerlick and P. von der Heyde, {\it Phys. Lett.} 
{\bf B 63} (1976) 446.}
\item{17.}{Y. Ne'eman and Dj. \v Sija\v cki, {\it Ann. Phys. (N.Y.)} {\bf 120} 
(1979) 292.}
\item{18.}{F.W. Hehl, J.D. McCrea, E.W. Mielke and Y. Ne'eman, {\it Phys. 
Reports} {\bf 258} (1995) 1.} 
\item{19.}{Dj. \v Sija\v cki and Y. Ne'eman, {\it J. Math. Phys.} {\bf 26} 
(1985) 2475.}
\item{20.}{Dj. \v Sija\v cki, "$\overline{SL}(n,R)$ Spinors for
Particles, Gravity and Superstrings", in {\it Spinors in  Physics and 
Geometry}, A. Trautman and G. Furlan eds. (World Scientific Pub., 1988),  
191-206.}
\item{21.}{V. Bargmann, {\it Ann. Math.} {\bf 48} (1947) 568.}
\item{22.}{A. Cant and Y. Ne'eman, {\it J. Math. Phys.} {\bf 26} (1985) 3180.} 
\item{23.}{Dj. \v Sija\v cki, {\it Phys. Lett.} {\bf B 109} (1982) 435.}
\item{24.}{Y. Ne'eman and Dj. \v Sija\v cki, {\it Phys. Lett.} {\bf B 200} 
(1988) 489.}
\item{25.}{Y. Ne'eman and Dj. \v Sija\v cki, {\it Phys. Rev.} {\bf D 37} 
(1988) 3267.}
\item{26.}{Dj. \v Sija\v cki and Y. Ne'eman, {\it Phys. Rev.} {\bf D 47} 
(1993) 4133.}
\item{27.}{S. Helgason, {\it Differential Geometry and Symmetric Spaces}
(Academic Press, New York and London, 1982).}
\item{28.}{T.E. Stewart, {\it Proc. Am. Math. Soc.} {\bf 11} (1960) 559.}
\item{29.}{N. Miljkovic, M. Sc. Thesis, Belgrade University (1987).}
\item{30.}{J. Mickelsson, {\it Commun. Math. Phys.} {\bf 88} (1983) 551.}
\item{31.}{A. B. Borisov, {\it J. Phys.} {\bf 11} (1978) 1057.}
\item{32.}{E. Eizenberg and Y. Ne'eman, {\it Membranes and Other Extendons}, 
(World Scientific Pub., Singapore 1995).}
\end